\documentclass[reprint,superscriptaddress]{revtex4-1}

\usepackage{graphicx}
\usepackage{amsmath}

\newcommand{\pmass}{m_{\gamma\prime}}

\begin{document}

\title{Cryogenic resonant microwave cavity searches for hidden sector photons}

\author{Stephen R. Parker}
\email{stephen.parker@uwa.edu.au}
\affiliation{School of Physics, The University of Western Australia, Crawley 6009, Australia}
\author{John G. Hartnett}
\affiliation{School of Physics, The University of Western Australia, Crawley 6009, Australia}
\affiliation{Institute of Photonics and Advanced Sensing, School of Chemistry and Physics, University of Adelaide, Adelaide 5005, Australia}
\author{Rhys G. Povey}
\affiliation{School of Physics, The University of Western Australia, Crawley 6009, Australia}
\affiliation{Department of Physics, University of Chicago, Chicago, IL 60637, USA}
\author{Michael E. Tobar}
\affiliation{School of Physics, The University of Western Australia, Crawley 6009, Australia}

\begin{abstract}
The hidden sector photon is a weakly interacting hypothetical particle with sub-eV mass that kinetically mixes with the photon. We describe a microwave frequency light shining through a wall experiment where a cryogenic resonant microwave cavity is used to try and detect photons that have passed through an impenetrable barrier, a process only possible via mixing with hidden sector photons. For a hidden sector photon mass of 53~$\mu$eV we limit the hidden photon kinetic mixing parameter $\chi~<~1.7\times10^{-7}$, which is an order of magnitude lower than previous bounds derived from cavity experiments in the same mass range. In addition, we use the cryogenic detector cavity to place new limits on the kinetic mixing parameter for hidden sector photons as a form of cold dark matter.
\end{abstract}


\maketitle

\section{Introduction}

Several theoretical extensions of the Standard Model introduce a hidden sector of particles that interact weakly with normal matter~\cite{sme1,sme2}. This interaction takes the form of spontaneous kinetic mixing between photons and hidden sector photons~\cite{paraphoton,holdem}. Paraphotons, hidden photons with sub-eV masses~\cite{paraphoton}, are classified as a type of Weakly Interacting Slim Particle (WISP)~\cite{wisps}. WISPs can also be formulated as compelling cold dark matter candidates~\cite{nelson2011,arias2012}. Indirect experimental detection of paraphotons is intrinsically difficult. The parameter space of kinetic paraphoton-photon mixing ($\chi$) as a function of possible paraphoton mass ($\pmass$) is extremely large, with many experiments and observations required to cover the relevant photon frequencies, ranging from below 1~Hz up to the optical regime. While solar observations strongly constrain hidden sector photon masses corresponding to higher optical frequencies~\cite{jaeckel2010}, the microwave region has yet to be fully explored.

One of the most sensitive laboratory-based tests to date is the light shining through a wall (LSW) experiment~\cite{exp1,exp2,exp3,exp4,exp5,exp6,exp7,povey2010,ADMX2010,betz2013}, whereby photons are generated on one side of an impenetrable barrier and then photon detection is attempted on the other side, presumably having crossed the barrier by mixing with paraphotons. In the microwave domain, mode-matched resonant microwave cavities can be used for the generation and detection of photons (emitter and detector cavity respectively)~\cite{Jaeckel08}. The low electrical losses of microwave cavities enables sub-photon regeneration~\cite{hartnett2011} and as such with appropriate experimental design extremely low levels of microwave power can be detected. Although other types of microwave cavity hidden photon searches have been developed~\cite{povey2011,parker2013}, they have yet to produce measurements that exceed the sensitivity of current LSW experiments. In this letter we discuss the design and results of a cryogenic LSW experiment and use the same setup to probe cold dark matter paraphoton / photon coupling.

\section{Experiment Design}

The sensitivity of a LSW microwave cavity experiment is dictated by~\cite{Jaeckel08}
\begin{equation}
\frac{\text{P}_{\text{DET}}}{\text{P}_{\text{EM}}}=\chi^{4}\text{Q}_{\text{DET}}\text{Q}_{\text{EM}}\left(\frac{\pmass c^{2}}{\hbar \omega_{\gamma}}\right)^{8}|G|^{2},
\label{eq:ptrans}
\end{equation}
where P$_{\text{DET}}$ and P$_{\text{EM}}$ is the level of power in the detecting and emitting cavity respectively, Q$_{\text{DET}}$ and Q$_{\text{EM}}$ are the cavity electrical quality factors, $\omega_{\gamma}$ is the photon / cavity resonance frequency and $G$ is a function that describes the two cavity fields, geometries and relative positions. Explicitly, $G$ is defined as
\begin{equation}
\begin{split}
G=k_{\gamma}^{2}\int\limits_{V_{\text{EM}}}d^{3}\mathbf{x}\int\limits_{V_{\text{DET}}}d^{3}\mathbf{y}\frac{\text{exp}\left(ik_{\gamma\prime}|\mathbf{x}-\mathbf{y}|\right)}{4\pi |\mathbf{x}-\mathbf{y}|} \\
\times A_{\text{EM}}\left(\mathbf{y}\right)\cdot A_{\text{DET}}\left(\mathbf{x}\right), \label{eq:gfac}
\end{split}
\end{equation}
with $A$ representing the normalized spatial component of the electromagnetic fields for the appropriate resonant cavity mode. The absolute value of $G$ is calculated as a function of $k_{\gamma\prime}$/$k_{\gamma}$, the paraphoton/photon wavenumber ratio. Calculation of Eq.~\ref{eq:gfac} is non-trivial and has previously been explored in detail~\cite{povey2010}. In this experiment we use the TM$_{0,2,0}$ resonant mode of two cylindrical cavities that are axially stacked and separated by 10~cm. 

Considering Eq.~\ref{eq:ptrans}, in order to maximize sensitivity to $\chi$ any LSW experiment should aim to minimize background power in the detector cavity and maximize power in the emitting cavity. The experiment should also use high Q cavities and optimize $G$ through appropriate cavity alignment and mode selection (using Eq.~\ref{eq:gfac}). As such, we operate the detector cavity cryogenically to reduce the level of thermal noise radiating from the cavity. Using a cavity made from niobium will also increase the Q factor as niobium is a type-II superconductor with a critical temperature of 9.2$^{\circ}$~K. In order to prevent power leakage between the cavities which is indistinguishable from a paraphoton signal~\cite{povey2010}, the emitter cavity is housed separately in a room temperature vacuum chamber.

\begin{figure}[t!]
\centering
\includegraphics[width=0.95\columnwidth]{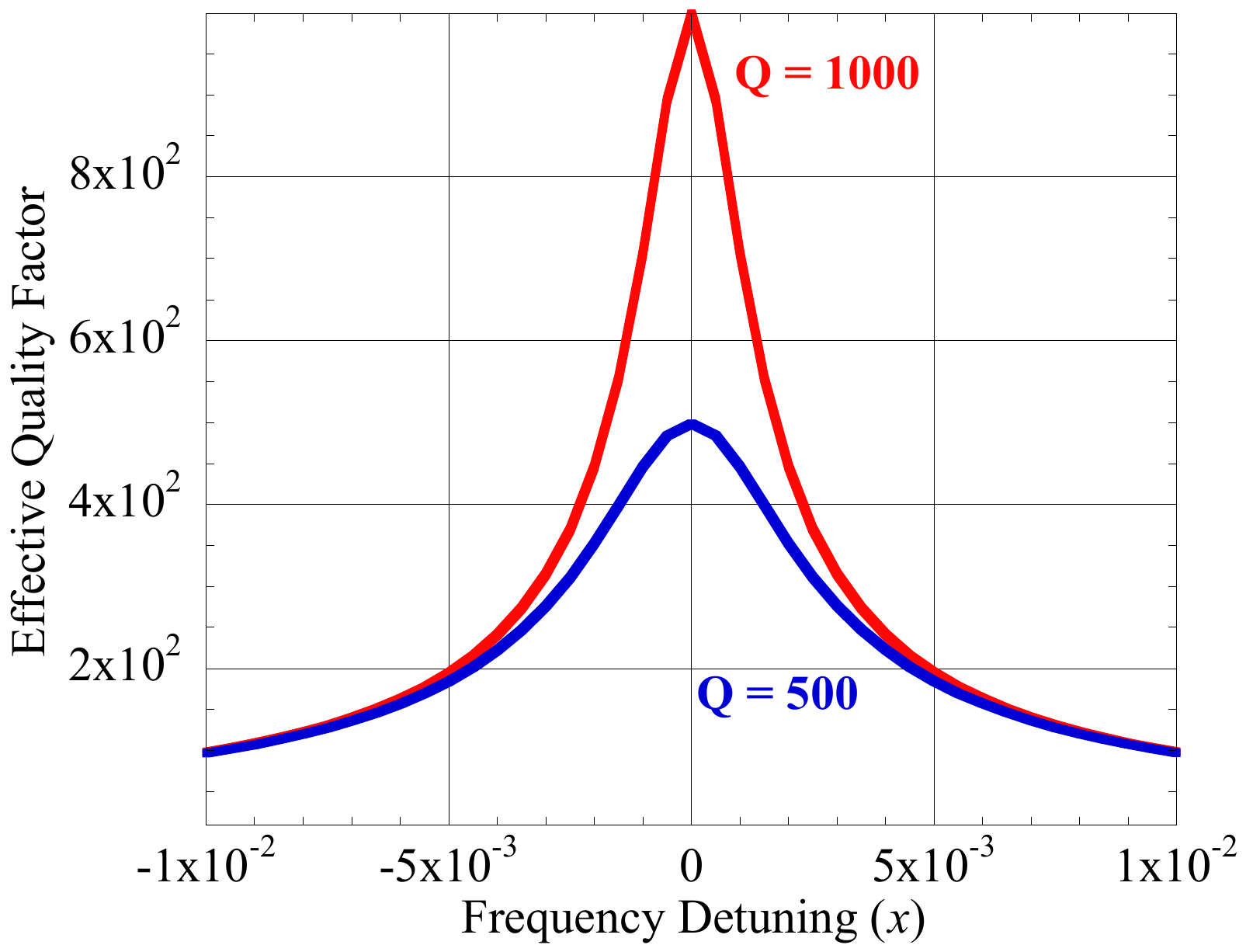}
\caption{(color online) Comparison of effective Q factor (Eq.~\ref{eq:Qeff}) as a function of frequency detuning for a central Q factor of 1000 (red curve) and 500 (blue curve).}
\label{fig:Qeff}
\end{figure}

Increasing the quality factor of the cavities will improve the sensitivity to $\chi$, but it will also reduce the cavity mode bandwidth making frequency matching between the emitter and detector cavities harder to obtain. It has been suggested that the optimal trade off is to use a high quality emitter cavity and a low quality detector cavity that has a large resonant mode bandwidth which could be easily tuned to overlap with the emitting mode~\cite{Jaeckel08}. When the cavities are tuned they have a common resonance frequency, $\omega_{0}$, as they become detuned the frequency shifts according to $\omega_{\text{CAV}}=\omega_{0}\left(1+\frac{x}{2}\right)$ where $x$ is the detuning parameter. The detuning of the cavities can be considered as an attenuation of the regenerated photon signal in the detector cavity, which can be expressed by defining a new mode with a central frequency at the detuned frequency and an effective Q factor that incorporates this attenuation,
\begin{equation}
Q^{\text{eff}}=\left|\frac{i Q \left(1+\frac{x}{2}\right)}{i+Q-Q\left(1+\frac{x}{2}\right)^{2}}\right|.
\label{eq:Qeff}
\end{equation}
Here the central frequency of the detector cavity is given by $\omega_{\text{EM}}\left(1+\frac{x}{2}\right)$. Mode-matching can be experimentally challenging but Fig.~\ref{fig:Qeff} explicitly demonstrates that there is no benefit to using a lower quality detector cavity as a higher quality cavity will always have a larger effective Q factor. Equation~\ref{eq:Qeff} should be combined with Eq.~\ref{eq:ptrans} to enable a more complete analysis of LSW experiments. Of course, one must always ensure that the cavities do not become detuned to the point of interacting with other resonant cavity modes.

\begin{figure}[t!]
\centering
\includegraphics[width=0.95\columnwidth]{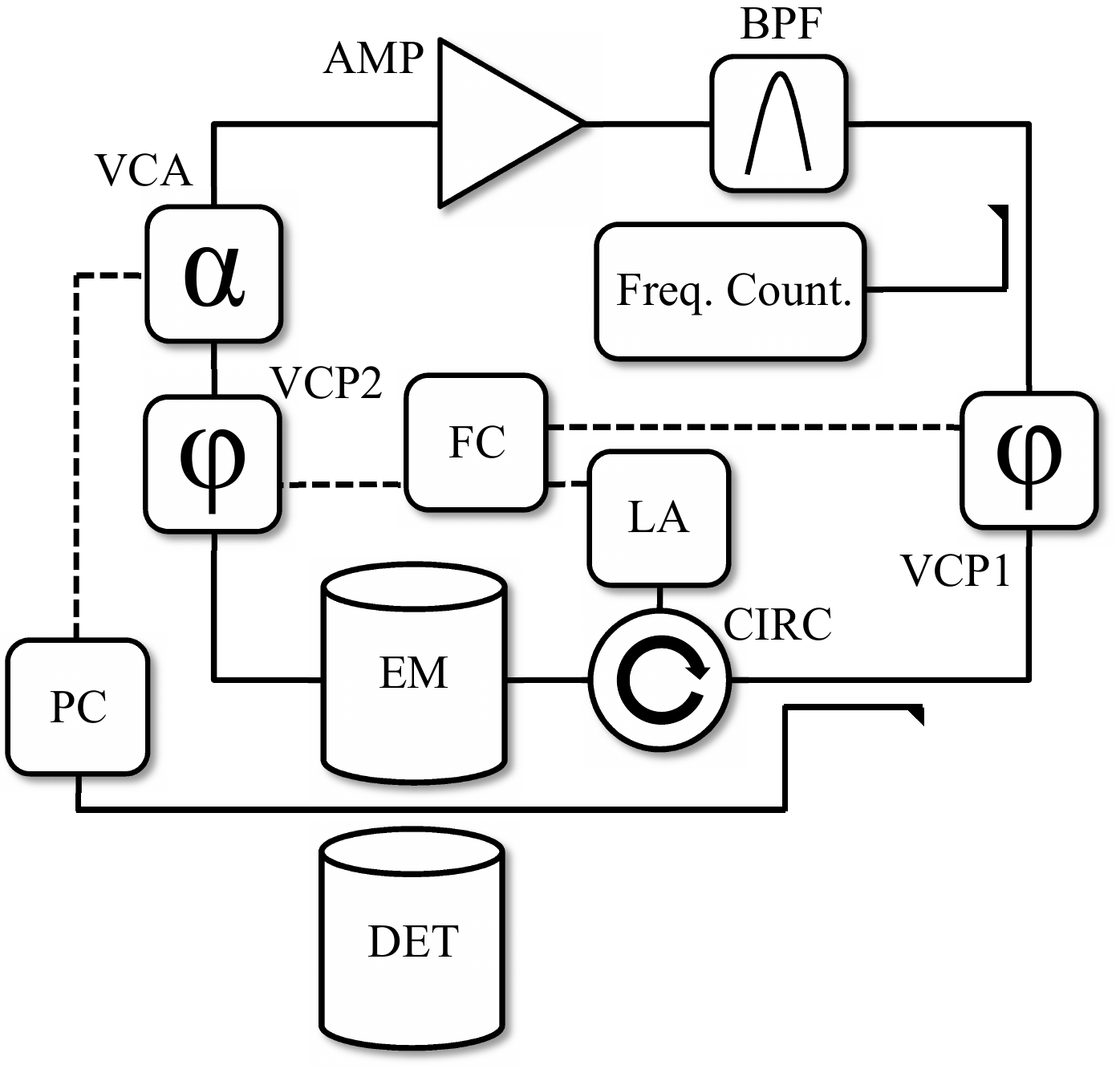}
\caption{Schematic of emitting cavity and control electronics. Components are labeled as follows: AMP = amplifier, BPF = band pass filter, VCA = voltage controlled attenuator, VCP = voltage controlled phase shifter, FC = custom frequency control electronics, LA = lock-in amplifier, CIRC = circulator and PC = custom power control electronics.}
\label{fig:em}
\end{figure}

A schematic of the emitting cavity and relevant electronics is shown in figure~\ref{fig:em}. The emitting cavity is a cylindrical copper cavity that is housed in a room temperature vacuum chamber to provide thermal isolation and minimize power leakage. When excited in the TM$_{0,2,0}$ resonant mode the Q factor was measured to be 3$\times$10$^{3}$ with a resonance frequency of 12.76~GHz. The cavity is anchored to a copper heatsink that is kept at a constant temperature via a Peltier temperature control feedback loop. The cavity acts as the frequency discriminating element of a microwave loop oscillator circuit, where a Pound phase locking scheme~\cite{pound1946} is employed to keep the signal stable and on resonance. A frequency counter referenced to a hydrogen Maser is used to track the resonance frequency of the cavity and then calculate the frequency detuning and effective Q factor of the detector cavity. The setpoint of the temperature control system can be adjusted to tune the resonance frequency of the cavity.

A power control system is used to keep the level of power in the cavity constant. Microwave power detectors are used to monitor the power incident on the cavity, the power reflected from the cavity and the power transmitted through the cavity. From this one can calculate the amount of power actually present in the cavity.

\begin{figure}[t]
\centering
\includegraphics[width=0.95\columnwidth]{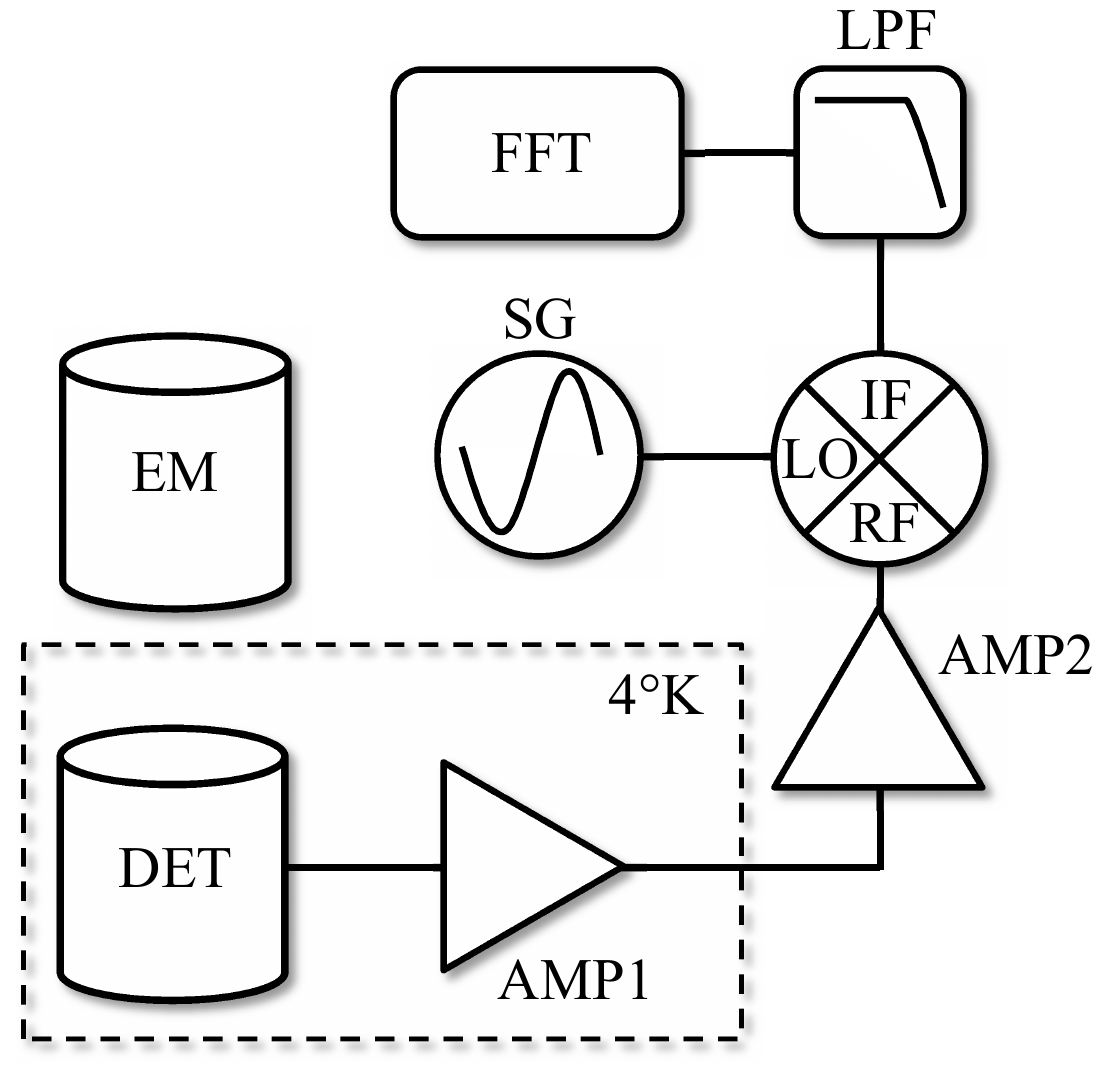}
\caption{Schematic of detector cavity and readout circuit. Components are labeled as follows: FFT = fast Fourier transform vector signal analyzer, LPF = low pass filter, SG = signal generator and AMP = amplifier. The dashed rectangle represents the cryogenic environment.}
\label{fig:det}
\end{figure}

Figure~\ref{fig:det} outlines the detector cavity and readout electronics (isolators are not shown). A superconducting niobium cavity is thermally anchored to the coldplate of a pulsed-tube cryostat system. A resistive heater is used to keep the temperature of the cavity stable at 5$^{\circ}$K. The Q factor of the TM$_{0,2,0}$ mode was measured as 9$\times$10$^{4}$. This value is considerably lower than previous work anticipated~\cite{patras2011}, which gave an estimate of $\sim$10$^8$. The reason for this discrepancy is that below the critical temperature the surface resistance of niobium is still limited by temperature~\cite{tesla2000}, which in turn limits the Q factor. To achieve higher Q factors on the order of 10$^{8}$ the cavity needs to be cooled below 2$^{\circ}$K and to have undergone stringent surface preparation procedures~\cite{NiCav}. With our current setup we were not able to cool the cavity below $\sim$4$^{\circ}$K.

A low noise HEMT amplifier~\cite{lnf} attached directly to the coldplate (approximately 4$^{\circ}$K) provides 31~dB of gain. The signal is amplified a second time at room temperature before being mixed with the output of a signal generator that is referenced to the same hydrogen Maser used to reference the frequency counter in the emitting circuit (Fig.~\ref{fig:em}). The signal generator is adjusted to give a mixer output with the signal of interest centered around approximately 1~MHz. The mixer produces a voltage signal proportional to the power incident on the RF port, which is then run through a Low Pass Filter (LPF) before being collected by a Fast Fourier Transform (FFT) vector signal analyzer.

The expected power spectrum of detector noise measured by the FFT can be calculated as follows. First we consider the transmission coefficient of the cavity,
\begin{equation}
\mathcal{T}=\frac{2\sqrt{\beta}}{\left(1+\beta\right)\left(1+2 i \text{Q}_{\text{DET}}\frac{\omega-\omega_{\text{DET}}}{\omega_{\text{DET}}}\right)},
\label{eq:transco}
\end{equation}
where $\beta$ is the coupling coefficient. Using Eq.~\ref{eq:transco} we can find the power spectrum of thermal noise emitted by the cavity combined with the noise contributions of the two amplifiers,
\begin{equation}
N_{\text{RF}}=\frac{k_{B}}{2}\left(\text{T}^{\text{C}}_{0}|\mathcal{T}|^{2}+\text{T}^{\text{A1}}_{\text{eff}}+\frac{\text{T}^{\text{A2}}_{\text{eff}}}{K_{\text{A1}}}\right)K_{\text{A1}}K_{\text{A2}},
\label{eq:nrf}
\end{equation}
where $k_{B}$ is Boltzmann's constant, T$^{\text{C}}_{0}$ is the physical temperature of the detector cavity, $\text{T}^{\text{A1}}_{\text{eff}}$ and $\text{T}^{\text{A2 }}_{\text{eff}}$ are the effective noise temperatures of amplifier 1 and 2 respectively (see Fig.~\ref{fig:det}) and $K_{\text{A1}}$ and $K_{\text{A2}}$ are the amplifier gains. The voltage spectrum measured by the FFT will be given by $S_{\text{MIX}}\sqrt{N_{\text{RF}}}$, where $S_{\text{MIX}}$ is the power to voltage conversion coefficient of the mixer, typically 10~V/$\sqrt{\text{W}}$. From Eq.~\ref{eq:nrf} it is clear that the gain of the cryogenic amplifier will render the noise contribution of the second amplifier insignificant. As such, the detection system will be limited by either the physical temperature of the cavity or the effective noise of the cryogenic amplifier.

\begin{figure}[t!]
\centering
\includegraphics[width=0.95\columnwidth]{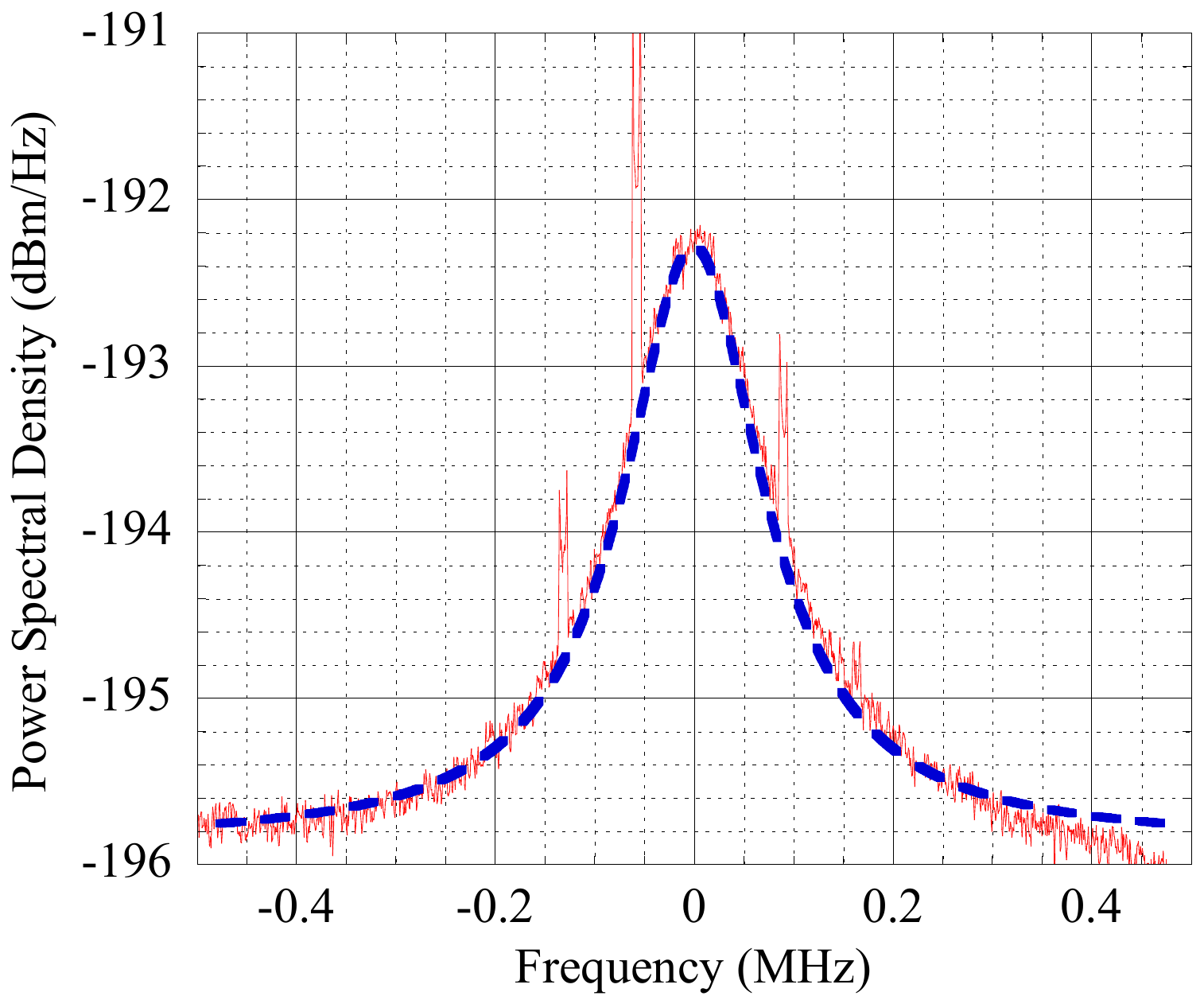}
\caption{(color online) Power spectral density of the the signal out of the detector cavity as a function of frequency offset from the resonance frequency of 12.76~GHz. The solid curve (red) is experimental data, the dashed curve (blue) is the expected profile calculated using Eq.~\ref{eq:nrf}}
\label{fig:psd}
\end{figure}

Determining the resonance frequency of the detector cavity can be achieved by observing the central peak of the noise spectrum measured on the FFT (see Eq.~\ref{eq:transco}) and noting the frequency of the signal generator driving the LO port of the mixer.

\section{Results}

The resonance frequency of the emitting cavity drifts by approximately 30~kHz every 24 hours, which is less than the bandwidth of either cavity and equivalent to a detuning factor of $x\approx$5$\times$10$^{-6}$. The temperature of the emitting cavity can be adjusted to return the resonance frequency to that of the detector cavity. The mean power in the emitting cavity during the same time period was 3.76~mW with a standard deviation of 0.6~$\mu$W.

Figure~\ref{fig:psd} shows the measured power spectral density of the detector cavity (red trace) compared to the expected spectrum (blue dashed trace) calculated from Eq.~\ref{eq:nrf}. The physical temperature of the detector cavity is 5$^{\circ}$K and the effective noise temperature of the cryogenic amplifier is $\sim$4$^{\circ}$K. The spikes that can be seen correspond to the 70~kHz modulation sidebands (and harmonics) generated by the lock-in amplifier as part of the Pound phase locked loop used for the frequency control of the emitting cavity. As there is no detectable signal at the resonance frequency of the emitting cavity, these spikes can be attributed to electronic leakage and not an authentic paraphoton signal. A true paraphoton signal would appear as a narrow excess of power at the same frequency as the emitting resonance.

The sensitivity of the experiment is limited by the thermal noise of the cavity (peaking at -192.2~dBm) and the effective thermal noise of the cryogenic amplifier. The difference in power between the two cavities is $\sim$198~dB, which is 80~dB lower than our previous experiment~\cite{povey2010}. As the values of the other factors in Eq.~\eqref{eq:ptrans} are similar, the sensitivity of our experiment to $\chi$ has been improved by 2 orders of magnitude.

\begin{figure}[t]
\centering
\includegraphics[width=0.95\columnwidth]{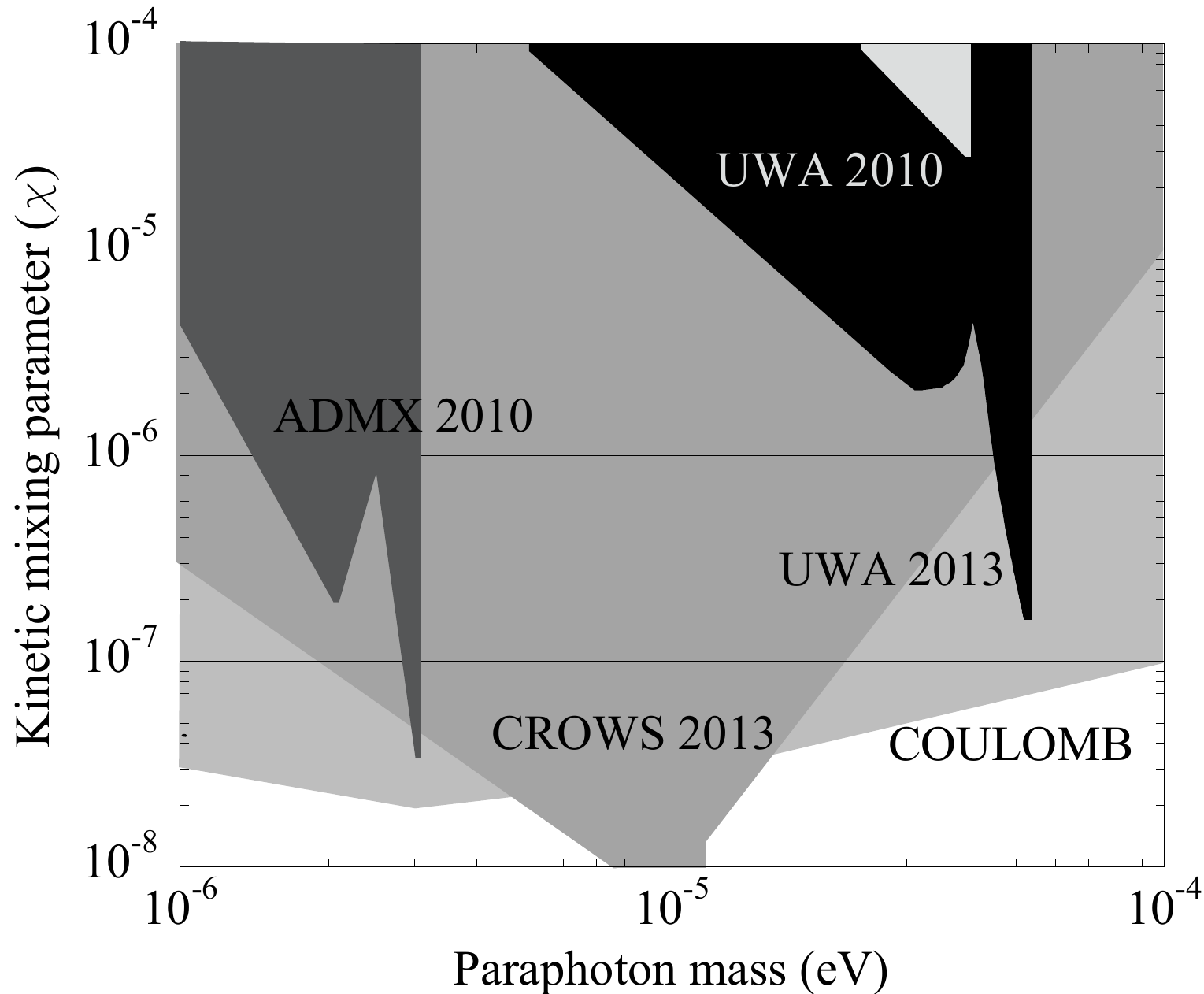}
\caption{Limits on the kinetic mixing parameter $\chi$ as a function of paraphoton mass. The mass range corresponds to frequencies from 240~MHz to 24~GHz. Different shaded regions correspond to bounds obtained by other experiments (refer to text for full description), with the bounds from this work presented in black.}
\label{fig:chi}
\end{figure}

Bounds for $\chi$ as a function of paraphoton mass are shown in Fig.~\ref{fig:chi}. The parameter space excluded by this experiment is shaded in black, with results from previous work~\cite{povey2010} shaded in light gray, bounds from the ADMX collaboration~\cite{ADMX2010} shaded in dark gray and new results from the CROWS experiment~\cite{betz2013} shaded in medium gray. Exisiting limits set by Coulomb law experiments~\cite{coulomb1,coulomb2} are also shown in light gray. For a paraphoton mass of 53~$\mu$eV we place the bound $\chi<1.7\times10^{-7}$, allowing us to exclude a significant region of the microwave frequency parameter space. These bounds are now comparable to the limits previously set by Coulomb law experiments~\cite{coulomb1,coulomb2} and the next generation of microwave cavity LSW searches will reach beyond this level of sensitivity. 

Areas for improving the experiment are clear. Cavity Q factors can be increased by several orders of magnitude by operating both cavities at lower temperatures to fully exploit the superconducting properties of niobium. Power levels in the emitting cavity can be further increased. Different cavity designs and modes can be explored, including the possibility of using tunable cavities to expand the area of parameter space the experiment is competitively sensitive to.

Resonant cavity experiments can also be used to set bounds on hidden sector photons as a form of Cold Dark Matter (CDM), hypothesized to exist via the misalignment mechanism~\cite{nelson2011,arias2012}. By turning off the emitting cavity of our experiment we are able to use our detector cavity to search for local CDM hidden photons. However, as our detector cavity can not be tuned we can only place bounds for particle masses falling within the bandwidth of our chosen resonant mode. Despite this, we are still able to probe uncharted parameter space that falls within the allowable region of CDM hidden photons. For this analysis we shall follow the work and assumptions of~\cite{arias2012}. For a single detector cavity the sensitivity to CDM hidden photons is given by
\begin{equation}
\text{P}_{\text{DET}}=\beta\chi^{2}\pmass\rho\text{Q}_{\text{DET}}\text{V}\mathcal{G},
\label{eq:cdmsens}
\end{equation}
where $\rho$ is the local density of CDM (typically assumed to be $\sim$~0.3~GeV/cm$^{3}$) and $\mathcal{G}$ is a dimensionless form factor similar to the axion microwave cavity haloscope form factor~\cite{hagmann1990},
\begin{equation}
\mathcal{G}=\frac{|\int d\text{V} \textbf{E}_{\text{DET}}\cdot\hat{\mathbf{n}}|^{2}}{\text{V}\int d\text{V} |\textbf{E}_{\text{DET}}|^{2}}.
\label{eq:cdmg}
\end{equation}
The unit vector $\hat{\mathbf{n}}$ is the direction of the CDM hidden photon field, which for now is taken to be the direction that optimizes the value of $\mathcal{G}$. As per~\cite{arias2012} we consider two scenarios regarding the orientation of the CDM hidden photon field. First we must multiply $\mathcal{G}$ by a factor of $\cos{\left(\theta\right)}^{2}$ to allow for different field directions. One possibility is that the CDM hidden photon field is homogeneous, although the direction is not known. By assuming that all directions are equally likely a value of $\cos{\left(\theta\right)}^{2}=0.0025$ is used to place conservative bounds on $\chi$. The other possibility is that the CDM hidden photon field is random and inhomogeneous so we average over all possible directions, meaning that $\langle\cos{\left(\theta\right)}^{2}\rangle=1/3$.

For our detector cavity operating in the TM$_{0,2,0}$ mode we use Eq.~\ref{eq:cdmg} to calculate a $\mathcal{G}$ value of 0.13. Using Eq.~\ref{eq:cdmsens} we place a limit on the kinetic mixing of 53~$\mu$eV CDM hidden photons of $\chi<6.14\times$10$^{-14}$ for a homogeneous CDM hidden photon field and $\chi<5.32\times$10$^{-15}$ for an inhomogeneous CDM hidden photon field. These values are over an order of magnitude lower than the estimated bounds presented in~\cite{arias2012} for the same hidden photon mass. Most importantly, this serves as a demonstration of the ability of microwave cavity experiments to reach unbounded and theoretically well motivated parameter space. With appropriate design considerations, future experiments will be able to search a wider range of CDM hidden photon masses and with a greater level of sensitivity.

\begin{acknowledgments}
The authors thank E.~N.~Ivanov and A.~Malagon for useful discussions. This work was supported by Australian Research Council grants DP130100205 and FL0992016.
\end{acknowledgments}

\end{document}